# Characterization of a DC glow discharge in $N_2$-$H_2$ with electrical measurements and neutral and ion mass spectrometry

**Audrey Chatain** [1,2,3,4], **Ana Sofia Morillo-Candas** [2,5], **Ludovic Vettier** [1], **Nathalie Carrasco** [1], **Guy Cernogora** [1], **and Olivier Guaitella** [2]

[1] LATMOS, Université Paris-Saclay, UVSQ, Sorbonne Université, CNRS, Guyancourt, France

[2] LPP, Ecole polytechnique, Sorbonne Université, Institut Polytechnique de Paris, CNRS, Palaiseau, France

[3] Corresponding author: audrey.chatain@latmos.ipsl.fr

[4] Present affiliation: Departamento de Física Aplicada, Escuela de Ingeniería de Bilbao, Universidad del País Vasco/Euskal Herriko Unibertsitatea (UPV/EHU), Bilbao, Spain

[5] Present affiliation: Paul Scherrer Institut, CH-5232 Villigen PSI, Switzerland

## Abstract

The addition of small amounts of $H_2$ were investigated in a DC glow discharge in $N_2$, at low pressure (~1 mbar) and low power (0.05 to 0.2 W.cm$^{-3}$). We quantified the electric field, the electron density, the ammonia production and the formation of positive ions for amounts of $H_2$ varying between 0 and 5%, pressure values between 0.5 and 4 mbar, and currents between 10 and 40 mA.

The addition of less than 1% $H_2$ has a strong effect on the $N_2$ plasma discharges. Hydrogen quenches the (higher) vibrational levels of $N_2$ and some of its highly energetic metastable states. This leads to the increase of the discharge electric field and consequently of the average electron energy. As a result, higher quantities of radical and excited species are suspected to be produced. The addition of hydrogen also leads to the formation of new species. In particular, ammonia and hydrogen-bearing ions have been observed: $N_2H^+$ and $NH_4^+$ being the major ones, and also $H_3^+$, $NH^+$, $NH_{2+}$, $NH_3^+$, $N_3H^+$ and $N_3H_3^+$.

The comparison to a radiofrequency capacitively coupled plasma (RF CCP) discharge in similar experimental conditions shows that both discharges led to similar observations. The study of $N_2$-$H_2$ discharges in the laboratory in the adequate ionization conditions then gives some insights on which plasma species made of nitrogen and hydrogen could be present in the ionosphere of Titan. Here, we identified some protonated ions, which are reactive species that could participate to the erosion of organic aerosols on Titan.

## 1- Introduction

$N_2$ and $H_2$ are very common gases, and they are often mixed up to create plasma discharges with efficient erosive properties. These properties are the core of many technological applications. As an example, $N_2$-$H_2$ discharges are used for nitriding, to harden metal surfaces [1,2]. The efficiency of $N_2$-$H_2$ plasmas to erode organic materials is also used in nuclear fusion, where $N_2$ is added to hydrogen plasma to inhibit organic film deposition on walls [3–5]. Another important aspect of $N_2$-$H_2$ plasmas is their capacity to synthesize of ammonia ($NH_3$). Surfaces play a key role in this production, which is currently closely investigated [6].

$N_2$-$H_2$ plasmas are also found in natural environments, especially in the ionospheres of some planets. In particular, the ionosphere of Titan – Saturn's largest moon – is a plasma of $N_2$, $CH_4$ and $H_2$. In more details, it is a dusty plasma and it includes a large number of solid organic aerosols [7]. Knowing the erosive properties of the $N_2$-$H_2$ plasma species, some works have investigated how the $N_2$-$H_2$ erosive component of Titan's ionospheric plasma could erode these organic aerosols [8,9]. These works use a DC glow discharge in $N_2$-$H_2$ to mimic the erosive part of Titan's ionospheric plasma and they expose analogues of Titan's organic aerosols to the plasma species. To interpret their results, an initial thorough characterization of the $N_2$-$H_2$ discharge is required. $N_2$-$H_2$ DC glow discharges have been previously studied by the community (e.g. [10–12]). However, none has been performed in the required experimental conditions of [9], at ~1 mbar and low power (0.05 to 0.2 W.cm$^{-3}$).

This work aims to characterize a DC glow $N_2$-$H_2$ discharge in this given range of working conditions. We analyze the electric field, the electron density, the ions and ammonia formation in the reactor. This work also aims to expand our current knowledge of $N_2$-$H_2$ glow discharges in additional plasma conditions.

The results are compared to a previous work [13] that characterized a radiofrequency capacitively coupled plasma (RF CCP) discharge in $N_2$-$H_2$ in similar working conditions and with the same analysis methods. We conclude on the interest to use such laboratory plasmas to simulate planetary ionospheres and get insights on the different species present, in particular in the case of Titan.

## 2- The instrumental setup

### 2.1- The reactor

DC plasma discharges are one of the simplest to create and study. Figure 1 shows a simplified scheme of the setup, and the different configurations used in this work.

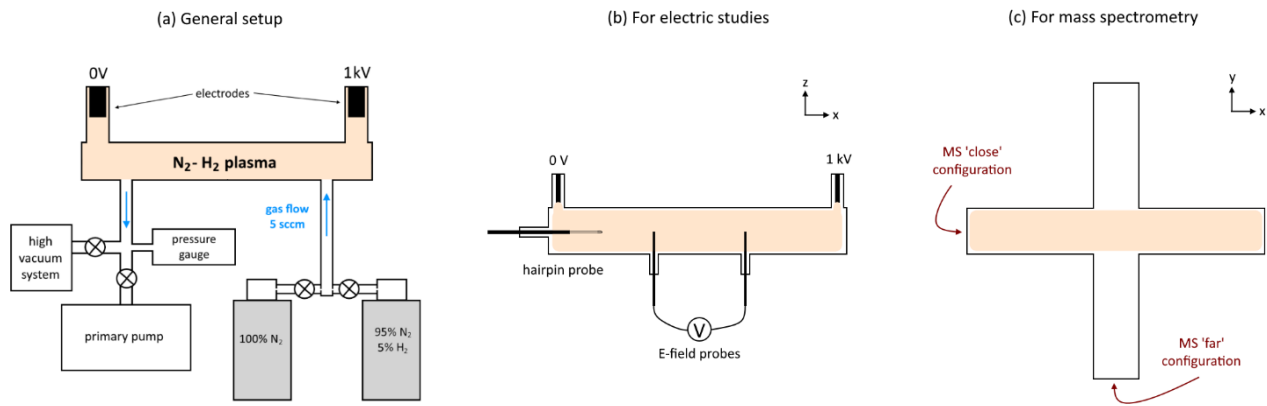

**Figure 1: Simplified schemes of the setup. (a) DC plasma reactor and gas circuit. (b) Configuration for electric studies: electric field probes and hairpin probe. (c) Configurations for the use of the mass spectrometer (MS).**

The reactor is a Pyrex tube 23 cm-long and with a 2 cm inner diameter. A gas mixture of $N_2$-$H_2$ at low density (0.5 to 5 mbar) is injected in the tube in continuous flow. We used high purity gases (> 99.999%). The gas from a 95% $N_2$ – 5% $H_2$ gas bottle (Air Liquide – CRYSTAL mixture) was diluted in pure $N_2$ (Air Liquide – alphagaz 2). The flow is fixed at 5 sccm (Standard Cubic Centimeters per Minute) by flowmeters from Brooks Instrument (range up to 5 sccm). The pressure is measured downstream of the reactor with two pressure gauges, one adapted for the mbar range (a capacitance gauge CMR 363 11mbar from Pfeiffer) and one for the ultra-high vacuum (a full range PKR250 from Pfeiffer). The gauges are situated downstream of the reactor to avoid any ignition of the plasma discharge on them. They were previously calibrated in different gases thanks to a third capacitance pressure gauge connected to a direct opening in the Pyrex tube.

Two cylindrical iron electrodes are positioned on both sides of the tube. They are coated with nickel with a small quantity of barium fluoride ($BaF_2$), an 'activator' whose role is to facilitate the extraction of electrons by ionic impact. A high voltage DC generator (maximum 3kV, 100 mA, regulated in current) creates a high voltage (1 to 3 KV) between the electrodes, leading to the ignition of the discharge inside the tube. The currents studied are 10-40 mA. The generator is coupled to a ceramic power resistor (10 kΩ) to stabilize the ignition of the discharge.

DC glow discharges are self-sustained discharges in which secondary electron emission from the cathode under ion bombardment compensates the loss of charges in the plasma. They have the advantage of developing with a peculiar structure well described in many text books [14], including a large volume homogeneous zone with relatively low and constant electric field called the positive column. This homogeneous positive column is filling the main part of the glass tube, which is a fundamental asset in the study of complex processes. The Pyrex walls confine closely the plasma, and could be a source of impurities. To limit such an influence, a great care is taken in cleaning the inside of the reactor. If previous experiments with the reactor involved organic compounds, the inside of the glass tube is first manually cleaned with ethanol – and sometimes diluted acetic acid. Then the tube is closed and pumped to high vacuum (~$10^{-6}$ mbar) with a turbo-molecular pump. To remove compounds – especially water – adsorbed on the walls, the tube is heated to ~80°C thanks to a heating cord and aluminum foil for 50 min. This removes most absorbed water, though we think that some is still present on the electrodes at this stage. Finally, the tube is let under high vacuum during > 30 min, until the mass spectrometer (further abbreviated 'MS') detects very few impurities (< 30 ct/s).

### 2.2- Setup for the electrical measurements

The electric field and the electron density are measured in a Pyrex tube reactor very similar to the one detailed just above and used in the MS section (2 cm diameter), but equipped with two tungsten pins separated by 20 cm at floating potential and a hairpin probe that is a type of microwave resonator (see Figure 1b). It is longer (55 cm), but this does not change the properties of the positive column in terms of electric field and electron density. Its wall temperature is controlled to 50°C.

### 2.3- Analysis of neutral and ionized gas species with mass spectrometry

Neutral molecules and positive ions are monitored with a quadrupole mass spectrometer (a Hiden analytical, electrostatic quadrupole plasma (EQP) series previously used in Chatain et al. [13]). The MS is connected to the reactor through a Pyrex junction of a few centimeters (see picture in Appendix 1 of Chatain et al. [9]). For the ion measurements, the collecting head of the MS required to be a few centimeters from the plasma, close to the grounded electrode (see the 'close' configuration in Figure 1c). However, to measure the perturbations that the collecting head could induce on the plasma by being close to it, neutral measurements have been performed in two configurations: close to the plasma (a few centimeters as for the ions) and far from the plasma (at ~30 cm, see the 'far' configuration in Figure 1c). To maximize the collection of newly formed species, the pumping system is always positioned in the same branch as the MS.

## 3- Electrical measurements and electron density

### 3.1- Method

The electric field of a glow discharge is constant in the positive column. Consequently, a simple measurement of the potential between the two tungsten probes in the plasma gives the electric field.

The electron density at the center of the discharge is measured by a hairpin resonator. This technique is described in Piejak et al. [15]. A tungsten wire shaped like the letter U (0.1 mm radius, tines 25 mm long, separated by 1.8 mm) is introduced in the plasma. It is attached to a glass tube containing a rigid coaxial cable terminated in a loop that inductively couples a small amplitude microwave signal to the probe. A microwave sweep oscillator (HP, 8350B) provides the microwave signal, repetitively swept over a small frequency range. The reflected amplitude is measured using a Shottky diode, and recorded with an oscilloscope (LeCroy, Waverunner LT584M). The resonant frequency is related to the electron density by the following equation:

$$n_e = \frac{\pi \cdot m_e}{e^2} \times \frac{f_r^2 - f_0^2}{\zeta} \qquad (1)$$

where $f_r$ is the resonant frequency with plasma, $f_0$ the resonant frequency without plasma, $m_e$ and $e$ the mass and charge of an electron. $\zeta$ is a correction factor to take into account the sheath forming around the probe, and the collisions in the gas phase induced at higher pressures [16]. This factor is equal to the product $\zeta = \zeta_s \cdot \zeta_c$, where $\zeta_s$ is the correction factor for the sheath and $\zeta_c$ is the collisions (i.e. pressure) correction factor. Both factors can be calculated according to the following expressions [16]:

$$\zeta_s = 1 - \frac{f_0^2}{f_r^2} \frac{\left[ ln\left(\frac{\omega - a}{\omega - b}\right) + ln\left(\frac{b}{a}\right) \right]}{ln\left(\frac{\omega - a}{a}\right)}$$

$$\zeta_c = \frac{1}{1 + \left(\frac{\nu_{en}}{2\pi f_0}\right)^2}$$

where $a$ is the hairpin wire radius, $\omega$ the distance between hairpins (from center to center), $b$ is the sheath radius and $\nu_{en}$ is the electron-neutral collision frequency. The sheath radius $b$ is calculated using the step-front sheath model from Piejak et al. [15]. The result of this fluid equations model can be fitted according to the following relation [17]:

$$\frac{b}{a} = 1 + 1.757 \left(\frac{a}{\lambda_D}\right)^{-0.63}$$

Where $\lambda_D$ is the Debye length, defined as $\lambda_D = \left(\frac{\varepsilon_0 k_B T_e}{n_e e^2}\right)^{1/2}$, with $\varepsilon_0$ the permittivity of free space, and $k_B$ the Boltzmann constant. As the sheath radius is dependent on the electron density itself through the definition of the Debye length, the system of equations is solved iteratively until the difference between the input and output $n_e$ value is less than 0.0001 %. The electron-neutral collision frequency $\nu_{en}$ was either taken as the tabulated value $\nu_{en}= 4.2 \cdot 10^9$ s$^{-1}$Torr$^{-1}$ [14] (used for the data plotted in Figure 2) or calculated with the method proposed by Peterson et al. [18]. The sheath correction was found to be dominant in similar experimental conditions (the positive column of a glow discharge) up to 1.5 Torr approximately. The collision correction increases rapidly with increasing pressure and becomes dominant at higher pressures (values up to 4 Torr were reported [16]). The uncertainty, however, also increases due to the sensitivity to the electron-neutral collision frequency value [18]. The differences between the electron densities obtained with the two $\nu_{en}$ values described above (tabulated or calculated) can be up to 40% for the highest pressure. The described procedure has previously been used to provide electron densities in a RF discharge in similar pressure conditions [19,20].

In each case, the total density of gas $N_g$ is estimated using the perfect gas law and gas temperatures adapted from measurements by Brovikova and Galiaskarov [21] and validated by a model by Pintassilgo and Guerra [22]. We took values at identical current density, and interpolated between current and pressure conditions. These measurements have been done in a pure $N_2$ plasma. As $H_2$ has a higher thermal conductivity than $N_2$, it could be possible that gas with a few percents of $H_2$ might be cooler than the pure $N_2$ case. However, a previous study in a similar discharge in $N_2$-$CH_4$ did not detect any variation of the gas temperature with the injected percentage of $CH_4$ in the range 0-5% [23]. As $CH_4$ dissociates partly in $H_2$ in the discharge, we can reasonably suppose that the gas temperature does not change significantly with the addition of a few percents of $H_2$ in the plasma.

### 3.2- Experimental results

Figure 2 shows the experimental results for varying experimental conditions. We observe that the electric field E decreases with the electric current, and that the reduced electric field E/$N_g$ decreases with pressure. Both E and E/$N_g$ increase with the addition of hydrogen to the gas mixture.

The electron density, and therefore the ionization degree, increases strongly with the electric current. In the case of pure $N_2$, at 1 mbar, the electron density increases from 1.5x10$^9$ cm$^{-3}$ at 10 mA to 7x10$^9$ cm$^{-3}$ at 40 mA. In these conditions, the ionization degree varies from 6x10$^{-8}$ to 4x10$^{-7}$.

The electron density and the ionization degree decrease with the addition of hydrogen. The difference induced by the hydrogen increases with pressure. In pure $N_2$ the electron density increases from 3.3 10$^9$ cm$^{-3}$ at 0.6 mbar to 3.9 10$^9$ cm$^{-3}$ at 4 mbar. On the opposite, with 5% $H_2$ the electron density decreases from 3.2 10$^9$ cm$^{-3}$ at 0.6 mbar to 1.8 10$^9$ cm$^{-3}$ at 4 mbar. We recommend to stay careful with the values at higher pressure (especially at 4 mbar) because the correction factor $\zeta$ to take into account the hairpin sheath and the collisions is more difficult to estimate after a few millibars.

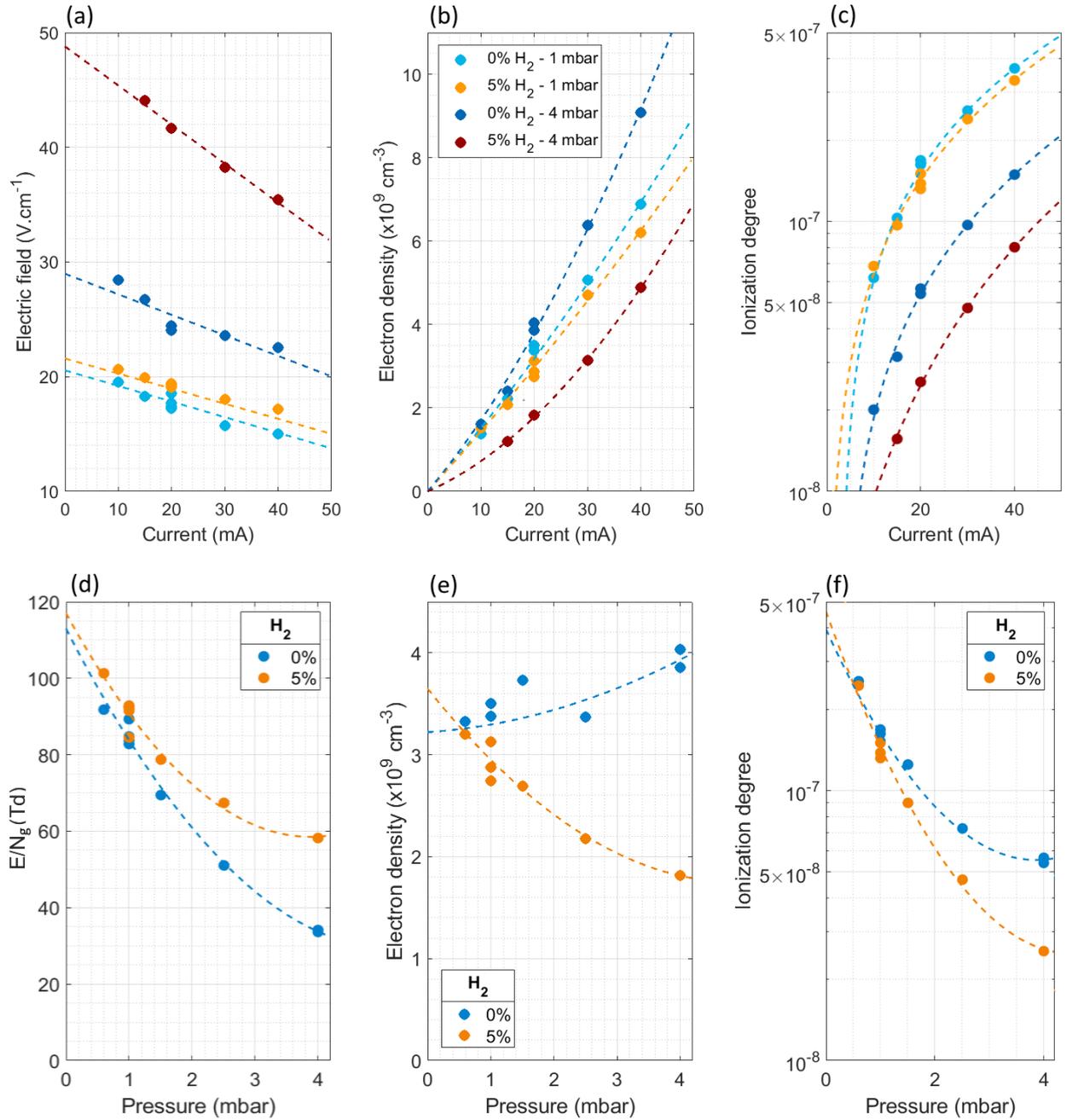

**Figure 2: (a) Electric field, (b) electron density and (c) ionization degree as function of the current. (d) E/N$_g$, (e) electron density and (f) ionization degree as function of the pressure at 20 mA. The electric field is measured by two floating potential probes, the electron density is measured by the hairpin probe and the total gas density N$_g$ is estimated using gas temperature measurements from Brovikova and Galiaskarov [21].**

### 3.2- Discussion of the results

As a point of comparison, the value for the electron density can be roughly estimated from the value of the electric field $E$ by the following equation, deduced from the definition of the current density $j$:

$$j = \frac{I}{\pi R^2} = q n_e \mu_e E \quad => \quad n_e = \frac{I}{q \mu_e E \pi R^2} \quad (2)$$

where $R$ is the radius of the discharge tube, $q$, $n_e$ and $\mu_e$ the charge, the density and the mobility of the electrons. $\mu_e$ depends on the $E/N_g$ ratio in the discharge. This is just a rough estimation because $E$, $n_e$ and $\mu_e$ are not constant over the tube section. For a typical case of 1 mbar and 20 mA, the gas temperature adapted from Brovikova and Galiaskarov [21] is 348 K. This gives a total density $N_g$ = 2.1x10$^{16}$ cm$^{-3}$, and then E/N$_g$ ≈ 86 Td. The electron mobility in these conditions is $\mu_e$ ≈ 5.2x10$^5$ cm².V$^{-1}$.s$^{-1}$ (Laplace database on LXCat - [24]). Finally, the estimation of the electron density in the pure N$_2$ plasma discharge (1 mbar, 20 mA) is $n_e$ ≈ 4.5x10$^9$ cm$^{-3}$, which is in the same order of magnitude than our measurement (3.4x10$^9$ cm$^{-3}$).

Besides, our observations are consistent with the measurements in the same conditions in a pure N$_2$ plasma by Cernogora et al. [25]. Their value for the electric field at 1 mbar and 20 mA is 17 V/cm, and the trends with pressure and current are the same as the ones observed in Figure 2. They measured the electron density by resonant cavity and found a value lower than in our work: 6x10$^8$ cm$^{-3}$ at 1 mbar and 20 mA. Nevertheless, the authors have doubts on this value, suspecting a mistake in the selection of the resonant mode (G. Cernogora, *private communication*). In any case, the trend with current and pressure is similar to the work presented here.

The increase of the electric field and the decrease of the electron density (Figure 2(a,b)) between pure N$_2$ and N$_2$-H$_2$ (0.95:0.05) is also observed by Thomaz, Amorim, and Souza [26]. The increase in the E field strength with the addition of small percentages of H$_2$ is an indirect effect already described in the literature [11,27,28]. It would be the consequence of the quenching on the one hand of the metastable states N$_2$(a'), N$_2$(A) and on the other hand of the upper vibrational states N$_2$(v) which contribute significantly to the ionization in N$_2$ plasmas by associative ionization processes producing N$_4^+$ then N$_2^+$. As soon as some H$_2$ is added, these processes are removed and the electric field must increase to compensate the ionization. The dominant ions become N$_2$H$^+$ and NH$_x^+$. Then, the increase in the electric fields leads to an increase of the electron drift velocity, that has to be compensated by a decrease of the electron density to keep the current constant.

### 3.3- Conclusions on the evolution of the N$_2$ discharge with the addition of H$_2$

Electric field and electron density measurements in our discharge are consistent with the glow discharge literature. We can therefore legitimately suppose that processes described in the literature also apply in our case [10,11,27–29]. Here are the main points. The addition of a few percents of H$_2$ leads to the quenching of the (higher) vibrational levels of N$_2$, as well as its energetic metastable states N$_2$(a') and N$_2$(A). In particular, these play a role in the ionization processes in a pure N$_2$ discharge. Consequently, E and E/N$_g$ increase with the amount of H$_2$ to increase the ionization efficiency (Figure 2(a,d)). Contrarily to E and E/N$_g$, the electron density decreases with the H$_2$ amount (Figure 2(b,e)). This leads to electrons less numerous, but more energetic that could possibly enable dissociation.

The increase of the electron energy increases slightly the dissociation of both N$_2$ and H$_2$ [29]. An important pathway to the formation of atomic H is *a priori* the quenching of N$_2$(a'): $N_2(a') + H_2 \rightarrow N_2 + 2\,H$. De Souza et al. [28] also showed that excited H atoms increase the dissociation of N$_2$ in N, with increasing H$_2$ percentage up to 2%. The dissociation rate in a pure N$_2$ discharge is ~1%. The vibrationally excited N$_2$(v) and metastable states are also present in the discharge (>0.005%). The quenching induced by H atoms with the addition of H$_2$ takes down the most energetic excited and metastable species. Nevertheless, the increase in the electron energy leads to an increase in the density of the lower energy excited states.

In conclusion, the chemical reactivity of the plasma increases with the addition of H$_2$. The main vector is the apparition of atomic H. But there is also an increase of atomic N with the addition of H$_2$ up to 2%, as well as an increase of the lower energy metastable species. Ions are also expected to evolve with the addition of H$_2$ (see Section 4.2).

## 4- Neutrals and positive ions by mass spectrometry

### 4.1- Neutral spectrum and quantification of ammonia

The discharge in N$_2$-H$_2$ leads to the formation of ammonia. Figure 3(a) shows the appearance of ammonia at the ignition of the discharge. m/z 17 can be due to NH$_3$, or to a fragment of H$_2$O, whose main peak is at m/z 18. According to the NIST database, the intensity of the fragment of H$_2$O at m/z 17, $I_{17}(H_2O)$, should be ≈ $0.21 \times I_{18}(H_2O)$. With $I_{17} \gg I_{18}$, the intensity $I_{17}$ after t = 19 min is mainly due to ammonia. A small increase of water is observed at the injection of gases, mostly because the pumping of water desorbed from the experiment's walls (in the reactor, and possibly also in the inox gas tubes between the reactor and the gas bottles) is less efficient at higher pressure. Air leaking from the outside of the reactor mostly contributes for $O_2$ at m/z 32. This $O_2$ impurity contribution is a bit lower than the desorption of water inside the experiment, which are both four orders of magnitude smaller than the main $N_2$ signal. An increase of water is also observed at the ignition of the plasma as more water is desorbed from the reactor electrodes. However, this contribution decreases in 10 minutes. Measurements presented in the following are done in the stabilized region, after the burst of desorbed water.

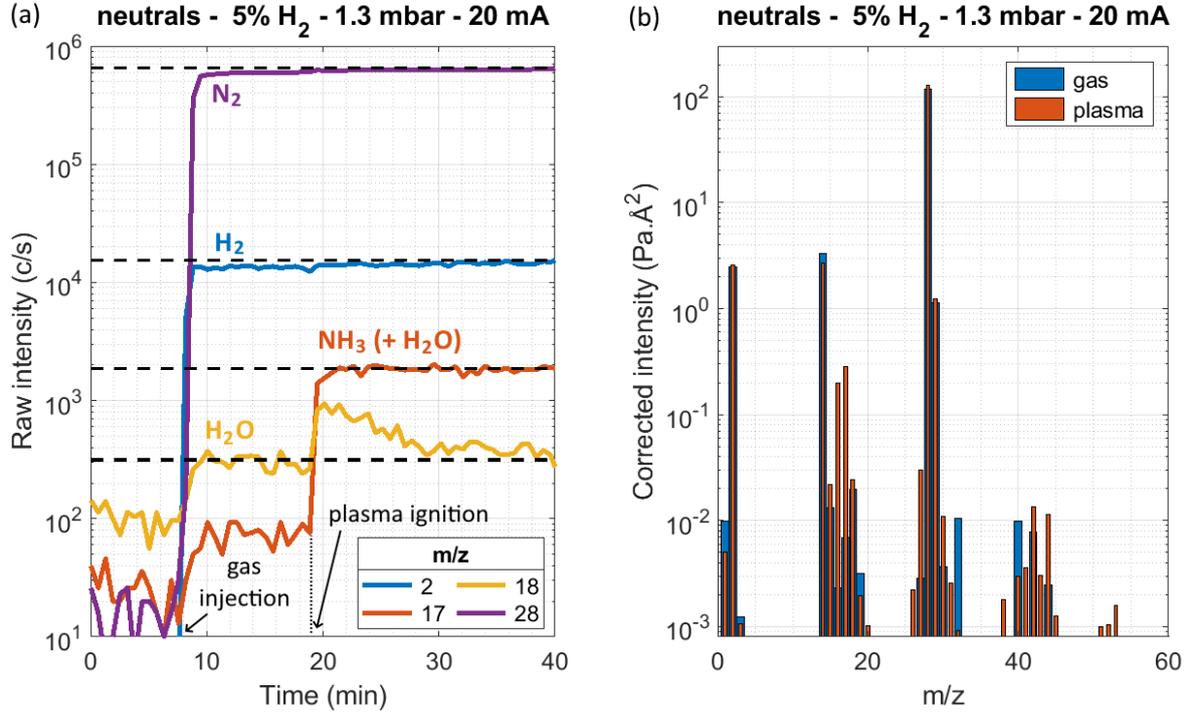

**Figure 3:** MS measurements at the ignition of a $N_2$-$H_2$ plasma discharge (5% $H_2$, 1.3 mbar, 20 mA). (a) Evolution in time of m/z 2 ($H_2$), 17 ($NH_3$ + $H_2O$), 18 ($H_2O$) and 28 ($N_2$). Gases are injected at t = 8 min, and the plasma is ignited at t = 19 min. (b) Total spectrum with only the gases, and after the ignition of the discharge.

Figure 3(b) gives the complete spectra before and after the ignition of the plasma. The appearance of ammonia is visible at m/z 16 and 17. No other major species are formed. One can notice that m/z 27 and 44 increase slightly; they are attributed to HCN and $CO_2$ formed by the carbon and water residues in the reactor.

To compare mass peak intensities in Figure 3(b), the intensity measured by the MS has been corrected by the MS transmission, which is mass-dependent. Details on the calibration are given in the Appendix D of Chatain et al. [13] that used the same mass spectrometer. However, as the calibration has not been done just before the measurement, a correction in amplitude is required to deduce the quantity of ammonia produced. For this purpose, we used the known partial pressure of $N_2$, $P_{N_2}$.

$$P_{NH_3} = \frac{I_{NH_3}}{\sigma_{NH_3}} \times \frac{\sigma_{N_2}}{I_{N_2}} \times P_{N_2} \qquad (3)$$

$\sigma_{NH_3}$ and $\sigma_{N_2}$ are the simple ionization cross sections of $NH_3$ and $N_2$ from impact of electrons at 70 eV in the MS ionization chamber. $\sigma_{N_2}$ is extracted from the Phelps database (=2.15+0.17 Å², www.lxcat.net/Phelps). $\sigma_{NH_3}$ is taken from the Hayashi database (=2.43 Å², www.lxcat.net/Hayashi). $I_{NH_3}$ and $I_{N_2}$ are the MS intensities at m/z 17 and 28 corrected from the MS transmission. The intensity at m/z 17 has also been corrected of the $H_2O$ contribution. In the reference plasma conditions shown in Figure 3, we obtain a partial pressure of ammonia at 0.23 ± 0.02 Pa, which gives a density of 5.0 ± 0.7 x$10^{13}$ cm$^{-3}$, and a percentage of 0.18 ± 0.01 %.

The ammonia detected in the reactor is sensitive to the plasma parameters. Figure 4 shows the influence of the $H_2$ amount, the pressure and the current values on the production of ammonia. Ammonia appears as soon as a small amount of hydrogen is injected in the discharge. However, no significant variation in the ammonia partial pressure is observed from 1 to 5% of $H_2$. We observe large variability on the data points for 3% $H_2$. The partial pressure of ammonia increases with the total pressure, but the mixing ratio of ammonia stays ~0.2% at all

- 7 -

pressures. The current has a strong impact on the production of ammonia. The partial pressure of ammonia produced is linear to the current value, with a slope of ~0.011 Pa/mA.

Similar results are obtained for the experiments with the MS being close or far from the plasma. This shows that the metallic head of the MS does not have a major influence on the $N_2$-$H_2$ plasma if positioned at a few centimeters from it. This validates the ion measurements performed in the 'close' configuration (in Section 4.2).

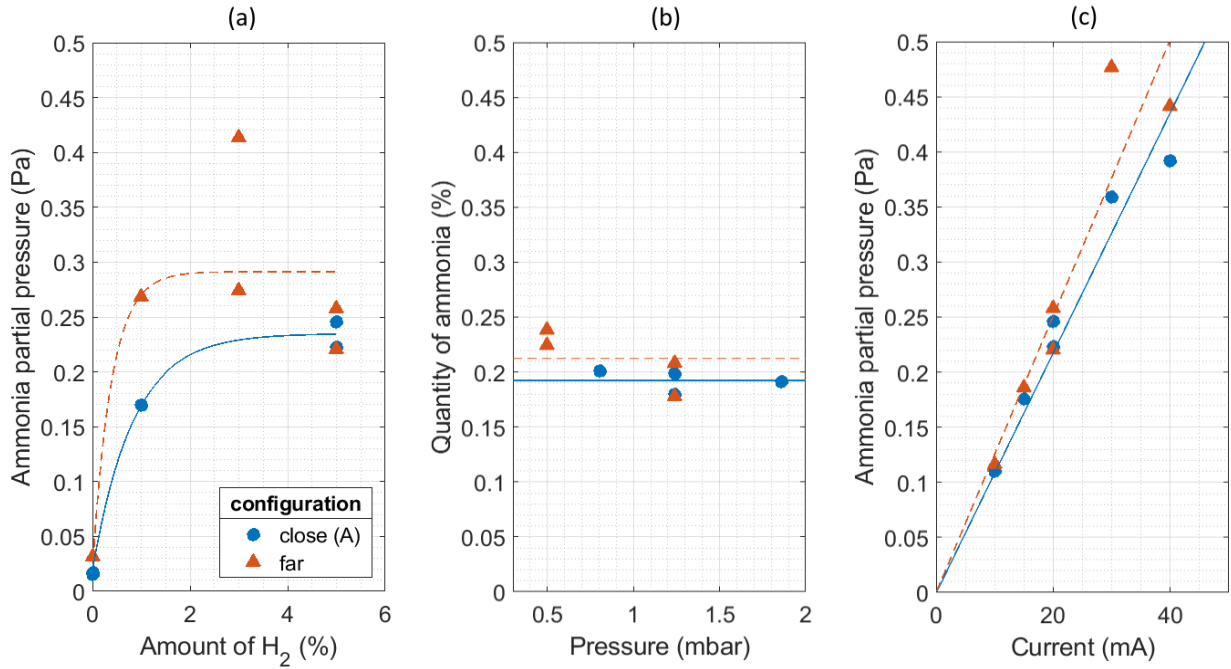

**Figure 4: Evolution of the ammonia quantity in the experiment for varying plasma parameters. The reference parameters are 5% $H_2$, 1.3 mbar and 20 mA. (a) Effect of the $H_2$ percentage on the ammonia partial pressure. (b) Effect of the pressure on the ammonia mixing ratio percentage in the gas phase. (c) Effect of the current on the ammonia partial pressure.**

The measured ammonia density is consistent with previous works. However, none have been performed in exactly the same conditions. Studies at 5% $H_2$, 2.7 mbar, 50 mA and 200 sccm measured an ammonia density of ~$1.10^{12}$ cm$^{-3}$ [29,30]. It is about one order of magnitude lower than in our experiment, which could be explained by their higher gas flow velocity (200 sccm compared to 5 sccm). Experiments at 90% $H_2$, 0.08 mbar, 150 mA in a tube of 10 cm in diameter give an ammonia production of ~1% [12]. This is about one order of magnitude higher than in our experiment, which can mainly be explained by the higher electric current used.

### 4.2- Detection of positive ions

Ions are directly measured by the MS, in the 'close' position. All measured ion spectra present the ion flux to the MS at a few centimeters from the plasma glow, these values are not exactly proportional to the amount of ions at the center of the discharge. The fluxes of ions larger than m/z 10 can be quantitatively compared (because the ions have a similar distribution in energy, see the appendix of [13]). However, $H_3^+$ at m/z 3 has a different energy distribution. Because of the need to choose a fixed value for the energy to probe with the MS, its intensity cannot be compared to other ions, only its relative variations with the plasma parameters give information. $H^+$ at m/z 1 is not a reliable measurement as it is at the limit of the MS mass range and certainly has a different energy distribution. First, 'background' measurements are acquired in a pure $N_2$ plasma. Figure 5(a) shows the reproducibility of the measurement, and Figure 5(b) presents the evolution of the spectrum with the addition of 5% hydrogen.

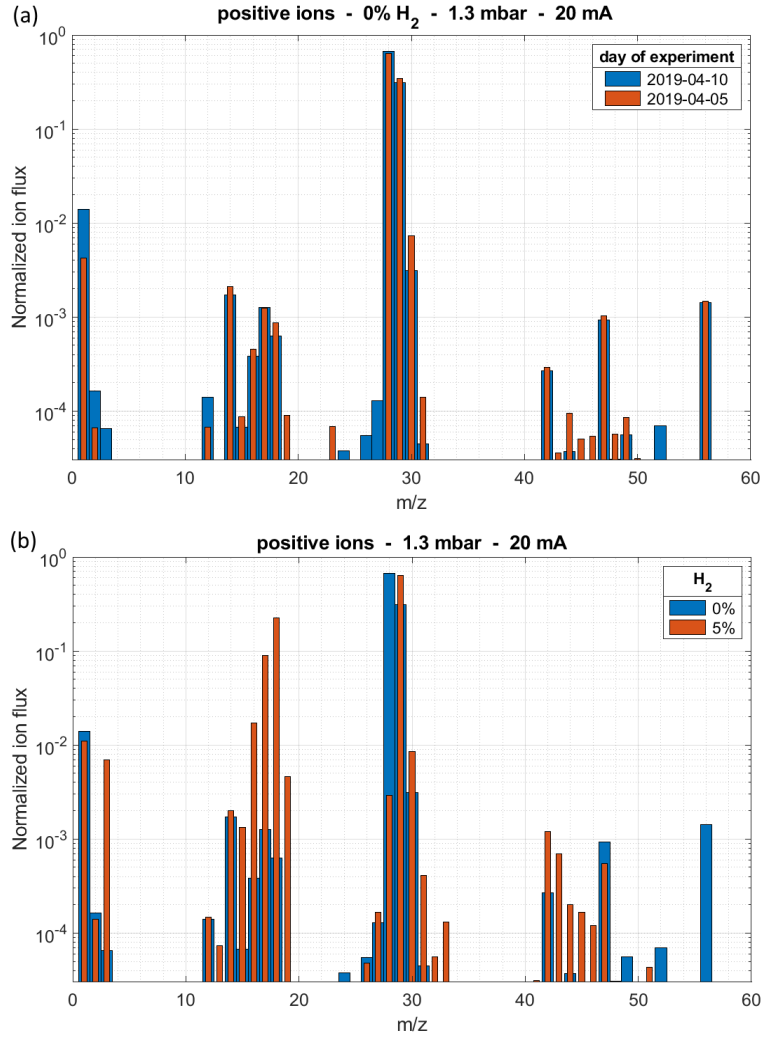

**Figure 5:** (a) Reproducibility of a positive ion mass spectrum in a pure $N_2$ plasma. (b) Comparison the $N_2$-$H_2$ plasma. Note: $H^+$ at m/z 1 and $H_3^+$ at m/z 3 should not be quantitatively compared to other ions. Their values are underestimated by the way the MS is configured.

With the addition of $H_2$ in the plasma, the ion species change. The main ion $N_2^+$ becomes $N_2H^+$ and ammonia ions $NH_x^+$ are formed in high quantities. One can also notice the appearance of $H_3^+$ and the increase of $H_3O^+$. The variations of the detected ions are presented in Table 1.

| m/z | in $N_2$ plasma | in $N_2$-$H_2$ plasma | $N_2$-$H_2$ case compared to pure $N_2$ | attribution |
|---|---|---|---|---|
| 3 | - | ++ | ↗↗↗ | **$H_3^+$** |
| 14 | + | + | - | $N^+$ |
| 15 | - | + | ↗↗ | **$NH^+$** |
| 16 | + | ++ | ↗↗ | **$NH_2^+$** |
| 17 | + | +++ | ↗↗ | **$NH_3^+$** / $HO^+$ |
| 18 | + | +++ | ↗↗↗ | **$NH_4^+$** / $H_2O^+$ |
| 19 | - | ++ | ↗↗↗ | $H_3O^+$ / $F^+$ (*) |
| 28 | +++ | ++ | ↘↘↘ | $N_2^+$ |
| 29 | +++ | +++ | ↗ | **$N_2H^+$** |
| 30 | ++ | ++ | ↗ | isotope $N_2H^+$ (at 0.7% of m/z 29)/ $N_2H_2^+$ |
| 31 | - | + | ↗ | $N_2H_3^+$ / $NOH^+$ |
| 42 | + | + | ↗ | **$N_3^+$** |

| 43 | - | + | ↗↗ | $N_3H^+$ |
| 47 | + | + | ↘ | $N_2F^+$ (*) |
| 56 | + | - | ↘↘ | **$N_4^+$** |

**Table 1: Positive ions peaks detected in $N_2$ and $N_2$-$H_2$ plasmas, and their suggested attributions. '+++' is for an ion flux > 3%, '++' for [0.3 ; 3]%, '+' for [0.03 ; 0.3]% and '-' for < 0.03%. (*) Some fluorine contamination is suspected at higher current conditions, triggered by the sputtering of $N_2^+$ or $N_2H^+$ ions on the cathode (containing $BaF_2$) or the PTFE (polytetrafluoroethylene) junction present close to the MS collecting head. This point is discussed in Chatain et al. [9].**

The variation of the ion fluxes with the injected amount of $H_2$, pressure and electric current are shown in Figure 6. With these variations in the plasma parameters, the balance between the four main ions changes strongly. $N_2^+$ (m/z 28) which is the main ion in pure $N_2$ plasma is hydrogenated to $N_2H^+$ (m/z 29) as soon as some hydrogen is added in the reactor (even from impurities). $N_2^+$ and $N_2H^+$ have isotopes at +1 m/z at 0.7%. Hydrogen also induces the formation of $NH_3^+$ (m/z 17) and $NH_4^+$ (m/z 18), respectively up to 8% and 12% at 1.3 mbar and 20 mA. As $NH_3$ is only 0.18% of the neutral gas molecules, we conclude the formation processes of $NH_3^+$ and $NH_4^+$ are extremely efficient. We note that $N_4^+$ is clearly visible in pure $N_2$ experiments, and completely disappears as soon as some $H_2$ is injected. A variation in pressure between 0.8 and 1.9 does not change greatly the ratios between the main ions. However, an increase of the electric current leads to a strong increase in the production of $NH_x^+$ ions, to the expense of $N_2H^+$. $N_2H^+$ decreases from 85% at 10 mA to 50% at 40 mA. On the opposite, $NH_4^+$ (resp. $NH_3^+$) increase from 5% (resp. 2%) to 30% (resp. 10%). This can be explained by the different dissociation cross sections of $H_2$ and $N_2$. $H_2$ is dissociated at lower energy than $N_2$, then $N_2H^+$ can be formed in the lower energy conditions. With increasing current, $N_2$ is more dissociated [25], increasing the formation of $NH_x^+$ ions to the expense of $N_2H^+$. $H_3^+$ ratio increases linearly with the injected amount of $H_2$, but decreases strongly with an increasing pressure or current. $H_3O^+$ also increases with the injection of $H_2$ as it favors the desorption of water from the walls by protonation. On the opposite to $H_3^+$, its production is strongly enhanced by an increasing electric current. m/z 42 to 45 also increase with the amount of $H_2$ and the electric current. The $H_2$ percentage has as greater effect on m/z 42 and 43 ($N_3^+$, $N_3H^+$), whereas the current has a stronger effect on m/z 44 and 45 ($N_3H_2^+/CO_2H^+$ and $N_3H_3^+$).

**Figure 6: Evolution of normalized ion fluxes with the injected amount of H$_2$, pressure and current. Dashed lines are added to guide the eye when appropriate (these are linear or 2$^{nd}$ degree polynomial fits). Note: H$_3^+$ at m/z 3**

**should not be quantitatively compared to other ions. Its values are underestimated by the way the MS is configured.**

We note that all the experiments presented in Figure 6 have been conducted jointly with experiments for another study [9]. For this purpose, an organic sample was positioned at ~30 cm from the plasma, to be inserted later in the plasma. Even if positioned far from the plasma, we suspect that the presence of the organic sample could change slightly the ion results. To test this possibility, an experiment in the reference conditions (5% $H_2$, 1.3 mbar, 20 mA) was conducted without organics in the vacuum chamber (presented in Figure 5). Results are mostly similar, with only a small increase in the ammonia ions $NH_x^+$ (x 2) and a decrease of $H_3^+$ (÷ 1.5) and of the trace carbon-containing ions at m/z 12 and 26-27-28 (÷ 4). So we can conclude that the presence of organics at a long distance from the plasma did not disturb significantly the plasma. The introduction of the sample inside the experiment but at long distance from the plasma does not bias the results.

# 5- Comparison to a $N_2$-$H_2$ RF CCP discharge and perspective to study Titan's ionosphere

Results with the DC discharge in $N_2$-$H_2$ are compared to measurements done on a radiofrequency capacitively coupled plasma (RF CCP) discharge in $N_2$-$H_2$ in similar experimental conditions and with the same mass spectrometer, previously published in Chatain et al. [13], abbreviated 'C20' below. The motivation of this work is that both discharges are used to simulate Titan's ionosphere [8,31]. Therefore, investigating the similarities and differences in the formed species in both discharges is fundamental to draw conclusions on Titan's case.

## 5.1- Electron density and the ionization ratio

The electron density measured in the RF CCP discharge (see Figure 8(a) in C20) and in the DC glow discharge (see Figure 2(b)) are similar, mostly between 1 x$10^9$ and 5x$10^9$ cm$^{-3}$. Conditions tested can reach lower values on the RF CCP discharge and higher values in the glow discharge because of the difference in power density (in average ~0.03 W.cm$^{-3}$ in the RF CCP discharge compared to ~0.12 W.cm$^{-3}$ in the DC discharge, see figure in Section 5.2). The comparison of the ionization degree obtained in the glow discharge and the RF CCP discharge (see Figure 8(b) in C20 and Figure 2(c)) gives very similar exploration ranges: between 2x$10^{-8}$ and 4x$10^{-7}$. For the two experiments, the ionization degree decreases with an increasing pressure.

How can these experiments be relevant for Titan's ionosphere? First, laboratory experiments are conducted at higher pressure than on Titan's ionosphere, but this is for several reasons. The very low pressures in Titan's ionosphere ($10^{-8}$ to $10^{-6}$ mbar) would require a huge reactor to ignite and limit the wall effects. Working at ~1 mbar is reasonable as it does not allow three-body reactions, similarly to Titan's ionospheric conditions. Besides, in such a dilute environment, collisions are rare, and the chemical evolution of the environment is slow. Increasing the pressure is therefore also a way to accelerate the processes to perform an experiment in hours or days instead of months or more. This point is further discussed in Chatain et al. [9]. In addition, the ionization degree is a quantity fundamental to reproduce a similar ion chemistry. The values reachable by the two experiments enable to probe the ionosphere of Titan between 900 and 1200 km (see Figure 7), at the altitude range of interest for studies investigating the formation and evolution of Titan's ionospheric aerosols [8,9].

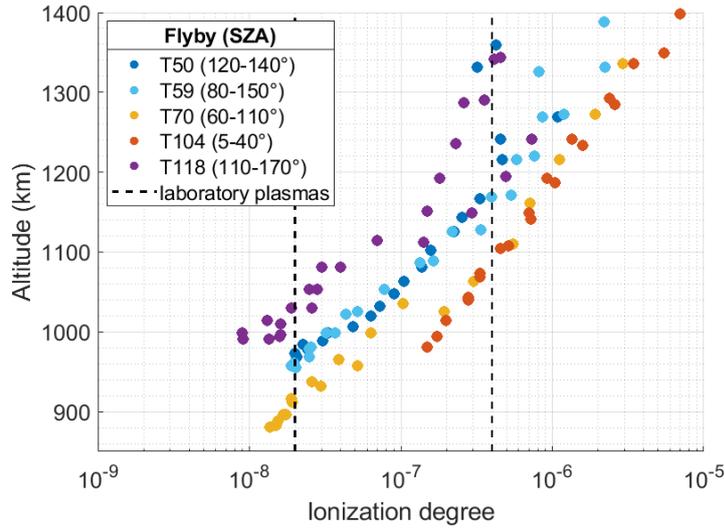

**Figure 7: Ionization degree (i.e. the electron density over total density ratio) as function of altitude on Titan for Cassini flybys at different Solar Zenith Angles (SZA). Data from Edberg et al. [32]. The ionization degrees reached in the laboratory are between the two dashed lines.**

## 5.2- Ammonia density

Ammonia is produced in the two experiments, RF CCP and DC, as soon as some hydrogen is injected in the gas phase and the plasma ignited. The reference conditions in both experiments are globally similar: (5% $H_2$, 0.91 mbar, 11 W, 425 $cm^3$) for the RF CCP discharge and (5% $H_2$, 1.3 mbar, 6 W, 55 $cm^3$) for the DC discharge. To investigate the evolution of ammonia with different parameters in the two experiments, Figure 8 compares the ammonia proportion. This enables to correct from the effect of pressure, which is a bit different in the two experiments. Indeed, as shown in Figure 8(b), the ammonia proportion is globally constant (~0.15-0.2%) with pressure in the DC discharge, and decreases in the RF CCP discharge. This difference is certainly due to the difference of geometry and material of the two reactors, that modifies the production of ammonia on the walls [33].

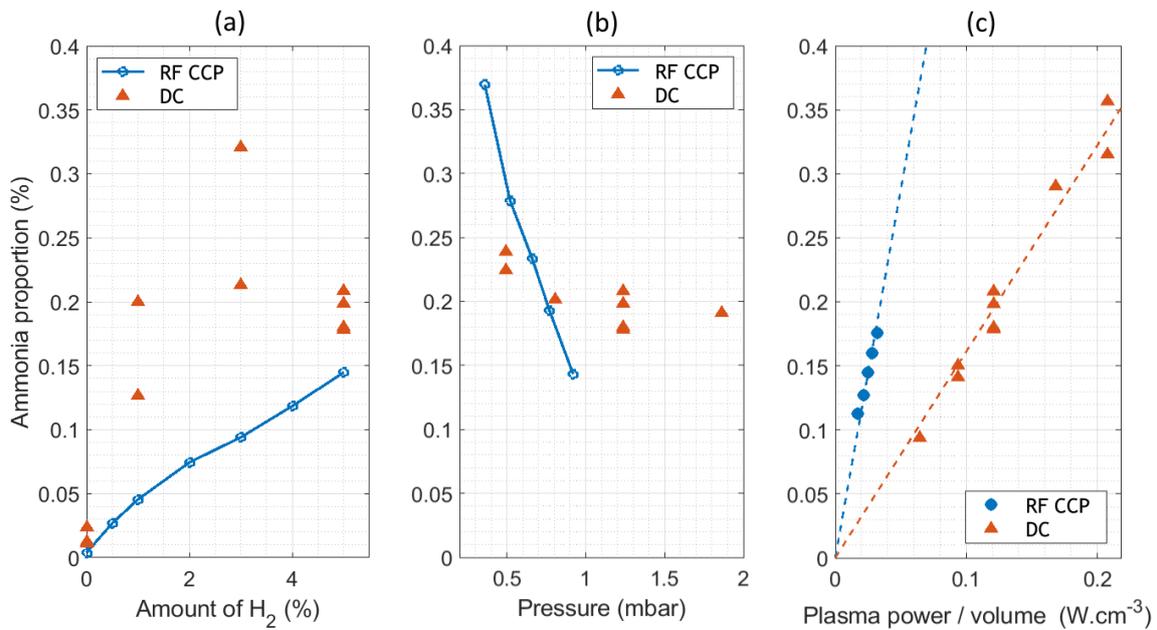

**Figure 8: Comparison of ammonia proportion detected in the $N_2$-$H_2$ RF CCP and DC discharges, for different plasma conditions. The reference condition for the RF CCP experiment is (5% $H_2$, 0.91 mbar, 11 W) and for the DC experiment (5% $H_2$, 1.3 mbar, 20 mA). In (c), dashed lines are linear fits of the data.**

Figure 8(c) shows that in both cases, ammonia is very sensitive to the power density given to the plasma and increases linearly with it. On the DC discharge, the power density in the positive column is computed from the electric field (E) measurement with:

$$\frac{Pw}{V} = \frac{E\left(\frac{V}{cm}\right) \times d(cm) \times I(A)}{\pi \times r^2 \times d \ (cm^3)} \qquad (4)$$

with r = 1 cm and d = 17 cm, the radius and the length of the positive column of the DC discharge. The power density is higher in DC in all the conditions studied because of the difference in volume $V_{RF\ CCP} \gg V_{DC}$. However, regarding the linear trends, for a given power density, more ammonia is formed in the RF CCP discharge compared to the DC discharge. This can be explained by the materials composing the walls of the two discharges: the metallic walls in the RF CCP discharge are more efficient to synthesize ammonia than the Pyrex walls of DC tube.

A different trend between the two experiments is observed with the injected $H_2$ amount (Figure 8(a)). A linear increase is observed in the case of RF CCP discharge, at the difference of a sharp increase and a plateau for the DC discharge. This also could be explained by the walls materials. Indeed, the formation of ammonia on metallic surfaces is quicker than on Pyrex [34–36]. In the case of the DC discharge, the surface is possibly saturated in $NH_3$ adsorbed precursors. Therefore, the limiting parameter in the production of ammonia is the available surface. In the case of the RF CCP discharge, the metallic surface does not saturate and the limiting parameter is the available $H_2$ amount.

In conclusion, in a $N_2$-$H_2$ plasma discharge, ammonia formation is enhanced by the addition of hydrogen and an increasing power delivered to the plasma. The production of ammonia depends on the surface material. These observations are difficult to extrapolate directly to Titan, especially because of the presence of walls, strongly catalyzing the production of ammonia, as shown in Jiménez-Redondo et al. [33]. However, it suggests the idea that the surface of Titan's aerosols could also have an influence on the production of ammonia. This point is studied further in Chatain et al. [9].

### 5.3- Positive ions

Positive ions play a fundamental role in the plasma chemistry, either in the DC discharge, the RF CCP discharge or on Titan. The different evolution of the normalized ion fluxes with the parameters of the discharge can be compared from Figure 6 (DC) and Figures 10-11-12 in C20 (CCP RF). We note that the given ion flux percentages underestimate the proportion of the light hydrogen ions.

#### *5.3.1- $H^+$, $H_2^+$, $H_3^+$ (m/z 1 to 3)*

These three hydrogen ions were found in the $N_2$-$H_2$ plasma discharges in both experiments. The MS measurements at m/z 1 ($H^+$) are not reliable, and the values obtained for m/z ≤ 3 are not comparable to other larger ions. However, we observed in both cases that $H_3^+$ was dominant over $H_2^+$. This can be explained by the quick conversion of $H_2^+$ into $H_3^+$ which is more stable. This result has been previously observed in laboratory plasmas and discussed in Méndez et al. [37] and Tanarro et al. [38].

As hydrogen is the major molecular species in the universe, its derived ions $H_x^+$ are found in many places at low density, such as the interstellar medium [39,40] and the upper atmospheres of giant planets [41,42]. Hydrogen is the third major species in Titan's ionosphere, after $N_2$ and $CH_4$ [43,44]. Therefore, its ions should be present and play a role in the ionospheric chemistry.

Hydrogen ions increase with increasing $H_2$ injected amount in both experiments. Similarly, as the percentage of $H_2$ increases with altitude, more hydrogen ions should be proportionally present at higher altitudes in the ionosphere of Titan. While the percentage of hydrogen ions is rather constant with pressure and power in the RF CCP discharge, it decreases in both cases in the DC discharge. This could be due to the certainly different electron energy distribution functions in the two discharges.

In particular, $H_3^+$ is relatively stable and is a strong protonating agent. Indeed, $H_2$ has a low proton affinity, and $H_3^+$ has high cross-sections relative to ion-molecule collisions. Milligan et al. [45] showed that in contact with species containing C, N and O, $H_3^+$ leads to the formation of new unsaturated molecules and ions in these low pressure environments. In the case of a $N_2$-$H_2$ plasma discharge, E. Carrasco et al. [46] observed that $H_3^+$ is a protonating agent leading to the formation of $N_2H^+$ from $N_2$, $NH_4^+$ from $NH_3$ and $NH^+$ and $NH_2^+$ from N.

*5.3.2- $N^+$, $NH^+$, $NH_2^+$, $NH_3^+$, $NH_4^+$  (m/z 14 to 18)*

In pure $N_2$ plasmas, only $N^+$ is present from the electron ionization of $N_2$ between m/z 14 and 18. Nevertheless, with the addition of hydrogen in the gas phase, the $NH_x^+$ ions are formed in quantity, in both the RF CCP and the DC discharges. For 5% injected $H_2$, at ~1 mbar and respectively 11W and 15 mA, the proportions of $NH_4^+$ and $NH_3^+$ in the two experiments are very similar, at ~8% for $NH_4^+$ and ~3% for $NH_3^+$. The evolution of the $NH^+$, $NH_2^+$ and $NH_3^+$ ions are exactly correlated with the variations of $NH_3$, with the $H_2$ injected amount, pressure and discharge power, for both the RF CCP and the DC discharges. Only $NH_4^+$ has a slightly different behavior in the case of the RF CCP discharge: its proportion increases with pressure instead of staying constant and it decreases with power instead of increasing.

The $NH_x^+$ ions are therefore strongly correlated to the quantity of ammonia produced in the discharge. Previous models of the chemistry in $N_2$-$H_2$ plasmas in DC glow discharges explain this correlation, as the ions are mainly directly formed from ammonia [38,46]. Indeed, $NH_4^+$ is produced by the protonation of ammonia by a protonating agent such as $H_3^+$ or $N_2H^+$. $NH_3^+$ is formed partly by direct electron ionization and by reaction with $H^+$ or $H_2^+$. Similar reactions can also lead to the production of $NH_2^+$ and $NH^+$ in smaller quantities.

On Titan, the detection of the $NH_x^+$ ions by mass spectrometry with INMS is difficult as the ions have masses superposed to the $CH_x^+$ numerous ions. In addition to the processes presented above in a $N_2$-$H_2$ plasma discharge, on Titan their production pathways can also include $CH_4$ [47]. $NH^+$, $NH_2^+$ and $NH_3^+$ are *a priori* quickly used to form other more stable molecules, as in $NH^+ + N_2 \rightarrow N_2H^+ + N$ [48]. Nevertheless, $NH_4^+$ stays strongly related to $NH_3$ density because the proton addition to $NH_3$ and the recombination of $NH_4^+$ are quick reactions. Therefore, on Titan $NH_4^+$ is often used as an indicator of the $NH_3$ density [49].

*5.3.3- $N_2^+$, $N_2H^+$  (m/z 28 to 31)*

In all the experiments except in pure $N_2$ plasma, $N_2H^+$ is the main ion, at 70-80%. Its variations with the hydrogen content and the pressure are the same for the two experimental setups. We observe a sharp increase with the addition of ~1% hydrogen in the gas phase and then a stabilization of its value. Concerning the pressure, the $N_2H^+$ proportion stays constant at least between 0.6 and 1.5 mbar. Nevertheless, in the case of a varying discharge power the results in the DC and RF CCP discharges are different. In DC, an increasing power (i.e. increasing current) leads to a large production of $NH_4^+$ (30% at 40 mA), which proportionally induces a decrease in the $N_2H^+$ percentage. As discussed in the above paragraph, it is not the case in the RF CCP discharge at the location of the measurement. Consequently, the $N_2H^+$ ratio stays relatively constant with an increasing plasma power in the RF CCP discharge. In the RF CCP discharge, $N_2H^+$ variations are always balanced by $N_2^+$, mainly because $NH_4^+$ is underestimated by the measurement in this non-homogeneous discharge. On the other hand, in the DC discharge, except in pure $N_2$ plasmas, the variations of $N_2H^+$ are balanced by $NH_4^+$ and $NH_3^+$.

In both discharges, the peak at m/z 30 is partly due to the $^{15}N^{14}NH^+$ isotope of $^{14}N_2H^+$, and maybe also to $N_2H_2^+$. The small peak appearing at m/z 31 in $N_2$-$H_2$ can be attributed to the diazenium ion ($N_2H_3^+$).

$N_2H^+$ is a common ion in ionized low pressure environments with nitrogen and hydrogen. It is mainly formed by the protonation of $N_2$ by $H_2^+$ or $H_3^+$ or by the reaction of $N_2^+$ with $H_2$ [38]. It does not react with $N_2$ or $H_2$. Then, it is found in the interstellar clouds [50,51].

On the opposite, $N_2H^+$ reacts easily with $CH_4$, as a protonating agent [38]. Besides, the detection of $N_2H^+$ on Titan INMS mass spectra is difficult as the major ion $C_2H_5^+$ also has a mass of 29 u. In any case, $N_2H^+$ plays an important role in Titan's complex chemistry [48].

*5.3.4- $N_3^+$, $N_4^+$  (m/z 42+ and 56)*

In both discharges, peaks at 42 and 56 u are observed in pure $N_2$ plasma. They are attributed to the $N_3^+$ and $N_4^+$ ions. According to Alves et al. [52] and Anicich et al. [53], $N_3^+$ and $N_4^+$ are formed by the reaction of $N^+$ and $N_2^+$ with $N_2$. Jiménez-Redondo et al. [33] also mention their production from $N_2$ excited states (Penning ionization). In any case, their peaks at 42 and 56 u would be hidden by other carbon-containing species in the ionosphere.

In the two experiments, $N_4^+$ disappears with the addition of $H_2$ to the gas mixture. It is consistent with the decrease of $N_2^+$, which is needed to form $N_4^+$. Nevertheless, $N_3^+$ stays present (as $N^+$), and other ions appear at m/z 43 to 46. The main one, at a mass of 43 u is certainly $N_3H^+$ [54–56]. 44 and 45 u could be due to $N_3H_x^+$ ions, but also to $CO_2^+$ and $CO_2H^+$ ions, formed from residues of $CO_2$ or $O_2$ in the reactor.

## 6- Conclusion

The addition of small amounts of $H_2$ has been investigated in a DC glow discharge in $N_2$. The plasma conditions are at low pressure (~1 mbar) and low power (0.05 to 0.2 W.cm$^{-3}$). The addition of less than 1% $H_2$ has a strong effect on the $N_2$ plasma discharges.

Electric field and electron density measurements in our discharge are consistent with the glow discharge literature, which allows us to interpret our observations in the light of previous works [11,27–29]. Hydrogen quenches the higher vibrational levels of $N_2$ and also some of its energetic metastable states. As a consequence, the discharge electric field increases, leading to the increase of the electron energy. Then, higher quantities of radical and excited species can be produced.

The addition of hydrogen leads to the formation of new species in the plasma. We quantified the production of ammonia in our discharge, which is consistent with previous works though none have been performed in exactly the same conditions. We also quantified the production of hydrogen-bearing ions for the first time in such experimental conditions. Most ions were expected because observed in previous works in $N_2$-$H_2$ discharges in different conditions [12]. In our case, the major ions measured are $N_2H^+$ and $NH_4^+$. Other ions observed are $H_3^+$, $NH^+$, $NH_2^+$, $NH_3^+$, $N_3H^+$ and $N_3H_3^+$.

The DC glow discharge has been compared to a RF CCP discharge in similar experimental conditions [13]. We observed some differences between the two, like a different trend of ammonia production with varying $H_2$ amount or pressure. Nevertheless, the main conclusion is that both discharges mostly led to similar observations. Such comparison of different $N_2$-$H_2$ discharges in the laboratory is motivated to give some insights on which $N_2$-$H_2$ plasma species could be present in the ionosphere of Titan, which has a similar ionization ratio. The fact that the two laboratory discharges show similarities in terms of ammonia and ion production supports the idea that both could be used to indirectly study processes happening in the ionosphere of Titan.

Titan's ionosphere also has a few percent of methane, which is the subject of a future inter-comparison study between the two discharges. In a plasma, methane dissociates in hydrogen and dihydrogen, which are species leading to the formation of protonated ions. This formation of protonated ions is the reason why we focused first on $N_2$-$H_2$ discharges: these ions are reactive species that could participate to the erosion of organic aerosols on Titan. This point is investigated in detail in Chatain et al. [9], a work done jointly with this paper.

## Data availability

Raw data from electric and mass spectrometry measurements, Matlab codes to process the data, processed Matlab variables, Matlab codes to plot the data and Matlab figures corresponding to the figures in the paper are available on Zenodo at the link: https://doi.org/10.5281/zenodo.7331908.

## Acknowledgments

ASMC was funded by LabEx Plas@par receiving financial aid from the French National Research Agency (ANR) under project SYCAMORE, reference ANR-16-CE06-0005-01. NC acknowledges the financial support of the European Research Council (ERC Starting Grant PRIMCHEM, Grant agreement no. 636829). AC acknowledges ENS Paris-Saclay Doctoral Program.